\documentclass[12pt]{article}
\usepackage[utf8]{inputenc}

\RequirePackage{calc}
\RequirePackage{natbib}
\RequirePackage[american]{babel}
\usepackage{chngcntr}

\usepackage{setspace}
\def\tblhead#1{\hline\\[-9pt]#1\\\hline\\[-9.75pt]}
\def\lastline{\\\hline}

\usepackage{rotating}
\usepackage{color}
\usepackage{url}
\RequirePackage{amsthm,multirow,amsmath,fancybox,amssymb}

\usepackage{graphics}

\makeatletter
  \def\title@font{\LARGE\bfseries}
  \let\ltx@maketitle\@maketitle
  \def\@maketitle{\bgroup%
    \let\ltx@title\@title%
    \def\@title{\resizebox{\textwidth}{!}{%
      \mbox{\title@font\ltx@title}%
    }}%
    \ltx@maketitle%
  \egroup}
\makeatother

\newcommand{\bx}{\mbox{\boldmath $x$}}
\newcommand{\by}{\mbox{\boldmath $y$}}

\newcommand{\bw}{\mbox{\boldmath $w$}}
\newcommand{\bQ}{\mbox{\boldmath $Q$}}

\newcommand{\bA}{\mbox{\boldmath $A$}}
\newcommand{\bD}{\mbox{\boldmath $D$}}
\newcommand{\btheta}{\mbox{\boldmath $\theta$}}
\newcommand{\bmu}{\mbox{\boldmath $\mu$}}

\newcommand{\bT}{\mbox{\boldmath $T$}}
\newcommand{\bSigma}{\mbox{\boldmath $\Sigma$}}
\newcommand{\bE}{\mbox{\boldmath $E$}}
\newcommand{\N}{\mbox{N}}
\newcommand{\E}{\mbox{E}}

\title{Pointless Continuous Spatial Surface Reconstruction}
\author{\vspace{-2mm}
Katherine Wilson$^1$ and Jon Wakefield$^{1,2}$\\
\vspace{-2mm}
{\small $^1$Department of Biostatistics, University of Washington, Seattle, USA}\\
{\small $^2$Department of Statistics, University of Washington, Seattle, USA}
}
\date{}

\begin{document}

\maketitle

\begin{abstract}
{The analysis of area-level aggregated summary data is common in many disciplines including epidemiology and the social sciences. Typically, Markov random field spatial models have been employed to acknowledge spatial dependence and allow data-driven smoothing. In this paper, we exploit recent theoretical and computational advances in continuous spatial modeling to carry out the reconstruction of an underlying continuous spatial surface. In particular,  we focus on models based on stochastic partial differential equations (SPDEs). We also consider the interesting case in which the aggregate data are supplemented with point data. We carry out Bayesian inference, and in the language of generalized linear mixed models, if the link is linear, an efficient implementation of the model is available via integrated nested Laplace approximations. For nonlinear links, we present two approaches:~a fully Bayesian implementation using a Hamiltonian Monte Carlo algorithm, and an empirical Bayes implementation, that is much faster, and is based on Laplace approximations.
We examine the properties of the approach using simulation, and then estimate  an underlying continuous risk surface for the classic Scottish lip cancer data.}
\\

\end{abstract}

\section{Introduction}
\label{sec1}

When modeling residual spatial dependence, it is appealing to reconstruct a continuous spatial surface, and this is the usual approach for point-referenced data.  Continuous reconstruction becomes more difficult when the data contain regional aggregates at varying spatial resolutions. In epidemiological studies, data is often aggregated for reporting or anonymization. While there exists a wealth of techniques to model regional data at a fixed resolution \citep{cressie:wikle:11,banerjee:etal:14}, these models do not extend in a straightforward fashion to situations where more than one resolution is used. In this paper, we develop methods for dealing with such situations.

We describe a number of motivating settings. The first scenario we consider is one in which data are collected from surveys at known locations and/or from censuses over large regions. Our interest in this problem arises from spatial modeling of demographic indicators in a developing world context. In many countries in this setting, demographic information is not available on all of the population, so data is collected via surveys, such as Demographic and Health Surveys \cite[DHS;][]{corsi:etal:12}. These surveys are typically stratified cluster designs with countries being stratified into coarse areas and into urban/rural, with enumeration areas (EAs) sampled within strata, and then households sampled within villages. In these surveys, the locations of the EAs, i.e.,~the GPS coordinates, are often available. We also consider census data, which is available at an aggregate level, e.g.,~ the average or sum of a variable over an administrative areal unit.
In the second scenario, we suppose
we have areal data only. In epidemiology and the social sciences this situation is the most common, since such data usually satisfy confidentiality constraints, and typically arises from aggregation over a disjoint, irregular partition of the study map, based on administrative boundaries. As an example, we consider incident lip cancer counts observed in 56 counties in Scotland over the years 1975--1980. These data provide a good test case, since they have been extensively analyzed in the literature; see \cite{wakefield:07} and the references there-in.  In this setting, we may view the continuous underlying surface as a device to induce a spatial prior for the areas that avoids the usual arbitrary element of defining neighbors over an irregular geography.
In each of these examples, we assume that there is a latent, continuous Gaussian random field (GRF) that varies in space, $\{S(\bx): \bx\in R \subset \mathbb{R}^2\}$ where $R$ is our study region of interest.

The situation with which we are concerned with in this paper is closely related to the change of support problem \cite[COSP;][]{gotway:young:02,cressie:wikle:11,gelfand:10,bradley:etal:16ACS}.
This problem occurs when one would like to make inference at a particular spatial resolution, but the data are available at another resolution. Much of this work focuses on normal data and kriging type approaches, with block kriging being used. For example, \cite{fuentes:raftery:05} combine point and aggregate pollution data, with the latter consisting of outputs from numerical models, produced over a gridded surface; MCMC is used for inference with block kriging integrals evaluated over a grid.
\cite{berrocaletal10} considered the same class of problem, but added a time dimension and used a regression model with coefficients that varied spatially to relate the observed data to the modeled output. 
\cite{moraga:etal:17} develop a similar framework to ours and use a stochastic partial different equation (SPDE) approach in order to relate two levels of pollution data. Specifically, the model they propose relates the continuous surface to the area (grid) level by taking an unweighted average of the surface at various points within each grid. We extend this work in several regards: most importantly, our model can accommodate non-normal outcomes and we also allow for a more complex relationship between the point-level process and the aggregated data. Therefore, we are able to address a wider range of situations.

\cite{diggle:etal:13} take a different approach for discrete data and model various applications using log-Gaussian Cox processes, including the reconstruction of a continuous spatial surface from aggregate data. Their approach is based on MCMC and follows \cite{li:etal:12LGCP} in simulating random locations of cases within areas, which is a computationally expensive step. 

A related problem to the COSP, is the modeling of data over time, based on areal data, but with boundary changes. \cite{lee:etal:17} analyze space-time data on male bladder cancer in Nova Scotia; the spatial aggregation changes over time, with the older data tending to be of aggregate form and the point data being the norm in more recent years. Building on previous work \citep{fan:etal:11,li:etal:12LGCP,nguyen:etal:12}, they use a local EM algorithm in conjunction with a local polynomial to model the risk surface.

We propose a three-stage Bayesian hierarchical model that can combine point and areal data by assuming a  common underlying smooth, continuous surface. We use the SPDE approach of \citet{Lindgren:etal:11} to model the latent field, which allows for computationally efficient inference. The paper is structured as follows. In Section \ref{sec:model} we describe the model and in Section \ref{sec:computation} the computational details. A simulation study in Section \ref{sec:simulation} considers a number of scenarios including: points data, areal data, and a combination of these data types. In Section \ref{sec:scotland} we illustrate the non-linear areal data only situation for the famous Scotland lip cancer example. Section \ref{sec:discussion} contains concluding remarks.

\section{Model Description}\label{sec:model}

We propose a general model framework for inference that can be used for data collected at points, over areas, or a combination of the two. We describe the model first for normal and then for Poisson data (as an illustration of a non-normal outcome), before concluding with a discussion of the model for the latent spatial surface.

\subsection{Normal Responses}

In general, models are specified at the point level. We describe the normal model in the context of modeling household wealth over a spatial region. Since we will  be concerned with observations at the area-level we will introduce general notation. The region of interest, $R$, is divided into $n$ disjoint areas denoted $R_i$, with $N_i$ households in area $R_i$, $i=1, \dots, n$. 
Let $Y_{ih}=Y(\bx_{ih})$ denote the $h$-th  response associated with location $\bx_{ih}$ (e.g., longitude and latitude), with covariate information $z_{ih}=z(\bx_{ih})$, $h=1,\dots,N_i$; we assume a single covariate only for notational simplicity, with the extension to multiple covariates being straightforward. 
The household-level model is $Y_{ih}~|~ \mu_{ih},~\sigma^2 \sim_{ind} \N(\mu_{ih}, \sigma^2)$, with $\mu_{ih} = \mu_{ih}(\bx_{ih}) = \beta_0 + \beta_1 z(\bx_{ih})  + S(\bx_{ih})$ and $S_{ih} = S(\bx_{ih})$ being the spatial random effect, where the spatial model is a GRF. 
The measurement error variance $\sigma^2$ is assumed constant (though this can easily be relaxed).
When data are available from a census we observe the average response in each of the areas $\bar{Y}_{i} = \frac{1}{N_i}\sum_{h=1}^{N_i} Y_{ih}$.
The induced area-level model is $\bar{Y}_{i} ~|~\mu_i,~\sigma^2 \sim \N(\mu_i, \sigma^2/N_i)$  where, 
 \begin{equation}
\mu_{i} = \frac{1}{N_i}\sum_{h=1}^{N_i}\{\beta_0 + \beta_1 z_{ih} + S_{ih}\} .\label{eq:norm1}
 \end{equation}
 
\subsection{Poisson Responses}

In the second case we consider, we assume that only the sum of all binary events, $Y_{i+} = \sum_{i=1}^{N_i}Y_{ij}$, is observed and recorded in area $R_i$.
The individual-level model is $Y_{ij}~|~p_{ij} \sim_{ind} \mbox{Bernoulli}(p_{ij}).$
We assume a rare event scenario, along with a log-linear model, so that, $p_{ij}=p_{ij}(\bx_{ij})=\exp\{\beta_0 + \beta_1z(\bx_{ij})+S(\bx_{ij})\}$.
We sum over all cases to give,
$
Y_{i+} ~|~ \mu_{i} \sim \mbox{Poisson}( \mu_i )
$, where,
 \begin{eqnarray}
\mu_{i} &=&  \sum_{j=1}^{N_i}\E [Y_{ij} | \bx_{ij}]\nonumber\\
&=& \sum_{j=1}^{N_i}
 \exp\{ \beta_0 + \beta_1 z(\bx_{ij} ) + S(\bx_{ij}) \}.
 \label{eq:pois1}
 \end{eqnarray}
If we have non-rare  outcomes and only observe the sum then the situation is far more difficult to deal with since the sum of binomials  with varying  probabilities is a convolution of binomials.
If we observe the individual outcomes $Y_{ij}$ (and not just the sum), then we can model each as binomial (i.e.,~we do not have to resort to the convolution).
The common situation in which disease counts and expected numbers are available across a set of areas is considered in Section \ref{sec:scotland}.

\subsection{Model for the Latent Process}

We assume a zero-mean latent, continuous GRF. There are many choices for describing how the form of the covariance changes with distance, but we follow \cite{stein:99} and others who make a strong argument for the Mat\'ern covariance function defined as,
\begin{equation}\label{eq:matern}
\mbox{Cov}[S(\bx_k),S(\bx_{k'})] = \frac{\lambda^22^{1-\nu}}{\Gamma(\nu)}(\kappa||\bx_{k} - \bx_{k'}||)^\nu K_\nu(\kappa||\bx_{k} - \bx_{k'}||),
\end{equation}
where $||\cdot||$ denotes Euclidean distance, $K_v$ is the modified Bessel function of the second kind and order $\nu$, $\kappa$ is a scaling parameter, and $\lambda^2$ is a variance parameter. In general, it is difficult to learn about the smoothness parameter $\nu$, and so it is conventional to fix this parameter; we follow this convention and set $\nu=1$. We define the practical range $\phi = \sqrt{8\nu}/\kappa$ as the distance at which the correlation drops to approximately 0.1.

\section{Computation}\label{sec:computation}

There are two steps to the computation, first the continuous latent surface is discretized in a convenient fashion (Section \ref{sec:latent}), and second the posterior is approximated. We begin with the normal case (Section \ref{sec:normalcomp}) before turning to the more difficult Poisson case (Section \ref{sec:poissoncomp}).

\subsection{Approximating the Latent Process}\label{sec:latent}

The major hurdle to the more widespread modeling of spatial data with a continuous surface has been the computation. In particular, inverting and finding the determinant of the Mat\'ern covariance matrix, which is in general not sparse, has been a roadblock when the number of points is not small.
However, recent work by \citet{Lindgren:etal:11} and \citet{simpson:etal:12b} detail the connection between GRFs and Gaussian Markov random fields (GMRFs). They first note that GRFs with a Mat\'ern covariance function are solutions to a particular stochastic partial differential equation (SPDE), and under certain relatively non-restrictive choices this produces a Markovian GRF (MGRF). 
They then show it is possible to obtain a representation of the solution to the SPDE using a GMRF.

We follow the SPDE approach and approximate the GRF over a triangulation of the domain (called the mesh) by a weighted sum of basis functions,
\begin{equation}\label{eq:SPDE1}
S(\bx) \approx \tilde{S}(\bx) = \sum_{k=1}^{m}w_k\psi_k(\bx),
\end{equation}
where $m$ is the number of mesh points in the triangulation, $\psi_k(\bx)$ is a basis function, and $\bw=[w_1,\dots,w_m]^\text{T}$ is a collection of weights. The distribution of the weights $\bw$ is jointly Gaussian with mean ${\bf 0}$ and sparse $m \times m$ precision matrix, $\bQ$, depending on spatial hyperparameters $\btheta = [\log \tau,~ \log \kappa]^\text{T}$ where $\tau^2 = 1/(4\pi \kappa^2\lambda^2)$; hence, $\bw$ is a GMRF. The form of $\bQ$ is chosen so that the resulting distribution for $\tilde{S}(\bx)$ approximates the distribution of the solution to the SPDE, and thus the form will depend on the basis functions. The basis functions are chosen to be  piecewise linear functions; that is, $\psi_k(\bx) = 1$ at the $k$-th vertex of the mesh and $\psi_k(\bx) = 0$ at all other vertices, $k=1,\dots,m$. This results in a set of pyramids, each with typically a six- or seven-sided base. Therefore, the spatial prior consists of  functions that are weighted linear combinations of these
pyramids, with the weights having a multivariate normal distribution. The sparsity of $\bQ$ eases computation.

For inference, the discretized version of the spatial prior is combined with the likelihood. In the setting where we have known locations, 
it follows from (\ref{eq:SPDE1}) that the value of the spatial random effect at an observation point, $\bx_{ij}$, can be approximated by a weighted average of the value of the GMRF on the three nearest mesh vertices. We can write, $S(\bx_{ij}) \approx \tilde{S}(\bx_{ij}) = \bA_{ij}^\text{T} \bw$, where $\bA_{ij}$ is an $m \times 1$-vector of weights that corresponds to the $ij$-th row of a sparse projection matrix $\bA$. The nonzero entries of $\bA_{ij}$, which correspond to the mesh points comprising the triangle containing $\bx_{ij}$, are proportional to the inverse distance from $\bx_{ij}$ to those mesh points, such that these values sum to one. In the case where the observation location, $\bx_{ij}$, is at a mesh vertex, $\bA_{ij}$ contains one non-zero entry that is equal to one.

For the normal response model when we have areal data, we use a fully Bayesian approach, since a fast computational strategy is available.
Specifically, the integrated nested Laplace approximation (INLA), an  approach for analyzing latent Gaussian models \citep{rue:etal:09}, can be used. INLA works by using a combination of Laplace approximations along with numerical integration to obtain approximations to the posterior marginals. The SPDE approach has also been implemented in the \texttt{R} package \texttt{R-INLA} \citep{lindgren:rue:15}, which allows for computationally efficient inference. For the Poisson response model, \texttt{R-INLA} cannot be used for data aggregated over areas; instead, we consider approaches that involve empirical Bayes (EB), MCMC, or a combination.

\subsection{Normal Responses}\label{sec:normalcomp}

The likelihood is normal with mean (\ref{eq:norm1}), and for simplicity we assume no covariates. The key to implementation is to approximate the integrated residual spatial area risk using the mesh. Defining $d(\bx_{ik})$ to be the ``relative'' population density at location $\bx_{ik}$ satisfying $d(\bx_{ik}) \geq 0$ and $\sum_{k=1}^{m_i} d(\bx_{ik}) = 1$, we obtain,
\begin{eqnarray}
\mu_i 
\approx \beta_0+\sum_{k=1}^{m_i}d(\bx_{ik}){w_{ik}} = \beta_0+\bD_i^\text{T} \bw,\label{eq:normA}
\end{eqnarray}
where $m_i$ is the number of mesh points in area $R_i$ and $\bD_i$ is an $m\times 1$ vector with up to $m_i$ nonzero entries $d(\bx_{ik})$.

This type of model can be fit using INLA, since $\bD_i^\text{T} \bw$ is Gaussian. See Appendix \ref{Ap:kenyasim1} for details on the implementation in the context of the simulation that we describe in Section \ref{sec:simulation}.
 
\subsection{Poisson Responses}\label{sec:poissoncomp}

For areal Poisson  data, we have the model $Y_{i+} ~|~ \mu_i \sim_{ind} \mbox{Poisson}(\mu_i)$.  We use a weighted average of the exponentiated spatial random effect at the mesh points contained in the area to form $\mu_i$. That is, and again ignoring covariates, we approximate the integral (\ref{eq:pois1}), to give 
\begin{eqnarray}
\mu_i 
\approx N_i \exp(\beta_0) \sum_{k=1}^{m_i}d(\bx_{ik})\exp(w_{ik}) = N_i \exp(\beta_0)\ \bD_i^\text{T} \bT \label{eq:poisA}
\end{eqnarray}
where $\bD_i$ is an $m \times 1$ vector as defined in Section \ref{sec:normalcomp} and $\bT=[\exp(w_1),\dots,\exp(w_m)]^\text{T}$.

Due to the structure of this model, it is not possible to use  INLA for fitting, but we describe three alternatives. First, a quick approximation is offered by EB with a Laplace approximation being used to integrate out the spatial random effects. To implement this, we use the \texttt{R} package \texttt{TMB} (which stands for Template Model Builder; \citealt{kristensen:14}). This is very efficient and estimates of the spatial hyperparameters and fixed effects can be computed within minutes. Second, we resort to Markov chain Monte Carlo (MCMC) methods. It is well known that in the Gaussian Process context, MCMC methods can be inefficient \citep{filippone:etal:13}. We opt to use a Hamiltonian Monte Carlo (HMC; Neal, 2011\nocite{neal:2011}) transition operator for updating $\bw$. Specifically, we first update the spatial hyperparameters $\btheta$ using a random walk proposal and then jointly update $\bw$ and $\beta_0$ using HMC. Finally, we consider a hybrid approach where estimates for the spatial hyperparameters $\btheta$ are found using the empirical Bayes approach and then, conditional on these estimates, posteriors for $\bw$ and any fixed effects are explored using MCMC methods. Details of these algorithms in the context of the Scotland example can be found in Appendix \ref{Ap:scotcomp}.

In both the simulation and the real data example, we use relatively vague priors; see Appendices \ref{Ap:kenyasim1} and \ref{Ap:scotcomp}.

\section{Simulation Study in the Normal Response Case}
\label{sec:simulation}

\subsection{Set Up}

We illustrate the method for normal responses via a simulation considering observations associated with points and observations associated with areas.
As a motivating example, we assume the aim is to construct a poverty surface; understanding the spatial structure of poverty and poverty-related factors is of considerable interest  \cite[e.g.,][]{gething:etal:15,minot:2005,okwi:2007}. Poverty has many different facets, and we take the wealth index as our measure \citep{dhswealth:2004}, which serves as a surrogate for long-term standard of living. We simulate a surface of the average wealth index within households. The wealth index is comprised of several variables such as household ownership of consumables, access to drinking water, and toilet facilities. The score is then standardized to have mean 0 and standard deviation 1. 

\begin{figure}[tbp]
\centering\includegraphics[width=1\linewidth]{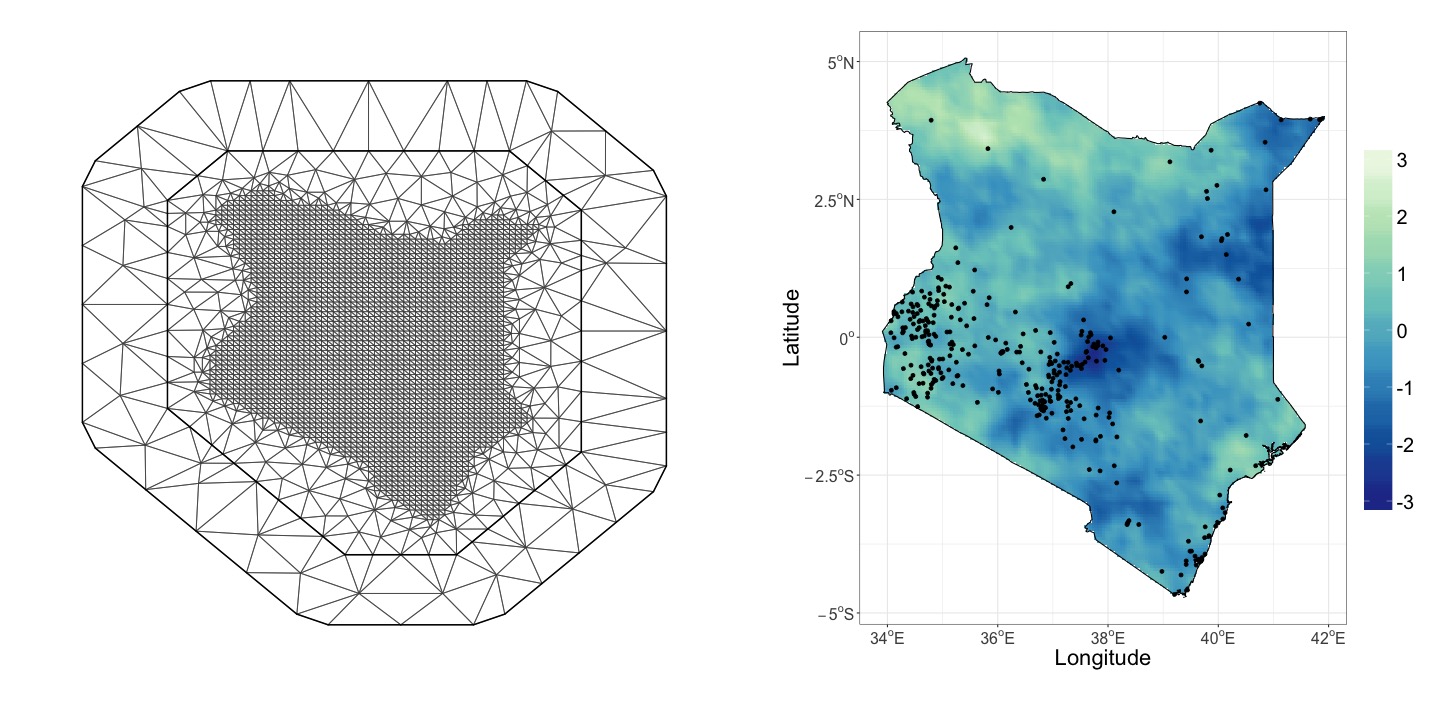}
\caption{Mesh (left) and latent spatial surface (right) used for the simulations. The mesh extends beyond the border of Kenya to avoid boundary effects. The black dots represent the locations of the 400 enumeration areas and black borders correspond to the boundaries of the 47 counties of Kenya.}
\label{Fig:kenyasetup}
\end{figure}

We will consider situations in which the wealth index is measured at point locations and we also consider incorporating census data, which provides the average wealth index at the area-level.
Observations associated with points are taken from a design that is informed by the Kenya DHS
\citep{KenyaDHS:15}. It is simplified in that we do not consider stratification or explicit cluster sampling for the 400 locations, which correspond to the centroids of enumeration areas (EAs) from the Kenya 2008 DHS. The dots on the plots in Figure \ref{Fig:kenyasetup} indicate the locations of these sampling points. We emphasize that these are point locations. 

Let $i=1,\dots,n$ index the administrative areas in Kenya and $j=1,\dots,n_i$ represent the EAs in area $R_i$. Hence, $\sum_{i=1}^n n_i=400$. Furthermore, let $h=1,\dots,N_i$ index the households included in the census in area $R_i$ and let $N_{ij}$ be the number of households surveyed at the $j$th location in $R_i$. For our simulation, the number of households participating in a survey, $N_{ij}$, ranges from 41 to 81, with mean 55 to give 21,496 households in total. We let $y_{ih}$ be the wealth index of household $h$ in area $R_i$.

We consider the data generating mechanism, $Y_{ih} ~|~ \mu_{ih},~ \sigma^2 \sim \N( \mu_{ih},\sigma^2)$ for $h=1,\dots,N_i$, $i=1,\dots,n$. Thus, for census data we assume the following model for the average wealth index in area $R_i$,
\begin{eqnarray}
\bar{Y}_{i} ~|~ \mu_{i},~ \sigma^2 &\sim& \N\left( \mu_{i},\frac{\sigma^2}{N_i}\right), \nonumber \\
\mu_i &=& \beta_0 + \frac{1}{N_i}\sum_{h=1}^{N_i}S(\bx_{ih}) \label{eq:censusmod}
\end{eqnarray}
where $S(\bx_{ih})$ is the value of the spatial random effect at geographic location $\bx_{ih}$. For survey data we assume the following model for the average household wealth index taken at EA $j$ in area $R_i$,
\begin{eqnarray}
\bar{Y}_{ij} ~|~ \mu_{ij},~ \sigma^2 &\sim& \N\left( \mu_{ij},\frac{\sigma^2}{N_{ij}}\right), \nonumber \\
\mu_{ij} &=& \beta_0 + S(\bx_{ij}) \label{eq:surveymod}
\end{eqnarray}
where $S(\bx_{ij})$ is the spatial random effect evaluated at the centroid, $\bx_{ij}$, of the EA. In this setting, $\sigma^2$ represents measurement error.

We assume that the spatial model is a MGRF with Mat\'ern covariance controlled by variance parameter $\lambda^2$ and scale parameter $\kappa$. Since the wealth index is standardized to have mean 0 and standard deviation 1, we have,
\begin{eqnarray*}
\E[Y_{ih}] &=& \E[\beta_0 + S(\bx_{ih})] = 0, \\
Var[Y_{ih}] &=& \E[\sigma^2] + Var[\beta_0 + S(\bx_{ih})] = 1.
\end{eqnarray*}
Therefore, we set $\beta_0 = 0$, and partition the variance as $\sigma^2 = 0.25$, and $\lambda^2 = 0.75$. To simulate the spatial surface, we use the SPDE approach, which requires a triangulated mesh. This mesh is shown in the left panel of Figure \ref{Fig:kenyasetup} with $m=2,785$ mesh points; these mesh points are approximately 15\ km apart in the interior of Kenya. We set $\kappa = \exp(1/2)$, which corresponds to a practical range of $\phi = \sqrt{8}/\kappa=1.72$ degrees. The simulated average household wealth index surface, i.e.,~~$\beta_0 + \tilde{S}(\bx)$, is shown in the right panel of Figure \ref{Fig:kenyasetup}, where the spatial effect $\tilde{S}(\bx)$ approximates $S(\bx)$.

To simulate data at the 400 EAs, we approximate (\ref{eq:surveymod}) by $\mu_{ij} = \beta_0 + \tilde{S}(\bx_{ij})$ where $\tilde{S}(\bx_{ij})$ is the simulated spatial effect at EA $j$ in area $R_i$. To simulate the census data we use gridded population estimates from SEDAC \citep{popdat}, which are available on a (approximately) 1\ km square grid at the equator. The gridded population estimates are then transformed to household estimates by dividing the population estimates by 3.9, the mean size of households in 2014 \citep{KenyaDHS:15}. We then approximate (\ref{eq:censusmod}) by $\mu_i = \beta_0 + \frac{1}{N_i}\sum_{g=1}^{G_i} N_{ig} \tilde{S}(\bx_{ig})$ where $N_{ig}$ is the household estimate for grid $g$ in area $R_i$, $N_i = \sum_{g=1}^{G_i}N_{ig}$ is the household estimate for area $R_i$, and $\tilde{S}(\bx_{ig})$ is the simulated spatial effect at the centroid of grid $g$.

We consider five different scenarios with varying levels of information available on location: (1) survey data only, (2) census data up to county level ($n=47$) only, (3) both survey data and census data up to county level, (4) census data up to provincial level ($n=8$) only, and (5) both survey data and census data up to provincial level. When we analyze survey and census data together, we assume the two data sources are independent, which in practice means that the surveyed population is only a small fraction of the total population.

To assess accuracy of the reconstruction under each scenario, we compute the mean squared error (MSE) and mean absolute error (MAE) of the spatial effect surface by
\[
\text{MSE} = \left(\sum_{i=1}^n m_i\right)^{-1}\sum_{i=1}^{n}\sum_{k=1}^{m_i}(\hat{w}_{ik}-w_{ik})^2, \qquad \qquad  \text{MAE} =  \left(\sum_{i=1}^n m_i\right)^{-1}\sum_{i=1}^{n}\sum_{k=1}^{m_i}\left|\hat{w}_{ik}-w_{ik}\right|
\]
respectively, where $\hat{w}_{ik}$ is the posterior mean and $w_{ik}$ is the ``true'' value of the spatial effect at mesh point $\bx_{ik}$.
Both the MSE and MAE are given in Table \ref{Table:sim} for all 5 scenarios. The top row of Figure \ref{Fig:kenyacomp} gives the sampling locations/areas, with each column corresponding to a different sampling scheme, and the middle and bottom rows the posterior means and standard deviations of the surface $S(\bx)$.

\subsection{Survey Data}
\label{sec:sim:points}

For our first simulation, we consider a situation in which we have survey data available from 400 EAs. To fit the model using \texttt{R-INLA}, we construct the projection matrix $\bA$ as described in Section \ref{sec:latent}. We fit  model (\ref{eq:surveymod}) using the SPDE approach. Computational details can be found in Appendix \ref{Ap:kenyasim1}.

\begin{sidewaysfigure}
\centering\includegraphics[width=\linewidth]{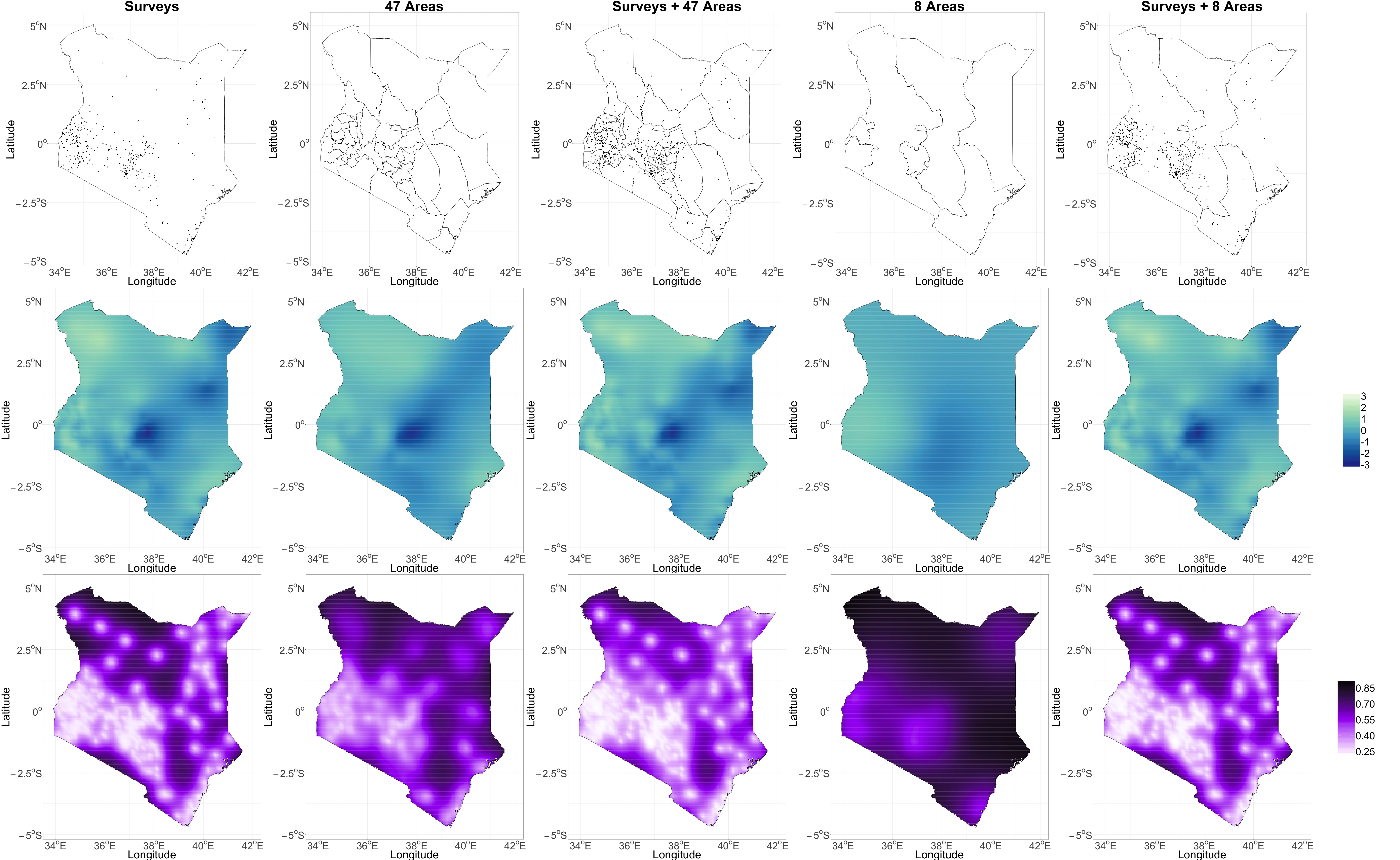}
\caption{Comparison of results under the five scenarios: 400 surveys with exact location (Surveys), census data at the county level (47 Areas), both survey and census data at the county level (Surveys + 47 Areas), census data at the provincial level (8 Areas), and both survey and census data at the provincial level (Surveys + 8 Areas). Top row is the data available under each scenario. Black dots are the locations of the enumeration areas. Middle row is the predicted (posterior mean) of the spatial surface. Bottom row is the posterior standard deviation of the predicted surface.}
\label{Fig:kenyacomp}
\end{sidewaysfigure}

Posterior medians and 95\% credible intervals (CIs) for the parameters are presented in Table \ref{Table:sim} and the predicted spatial random effect surface is depicted in Figure \ref{Fig:kenyacomp} (left column). In general, the posterior medians are relatively close to their true values and all credible intervals cover the true value, though are fairly wide. The predicted spatial surface (posterior mean) over Kenya is visually similar to the true spatial surface, though there is some attenuation. Regions of Kenya that have a higher spatial effect are predicted to be lower and vice versa; this shrinkage to the mean phenomenon is well known in the spatial literature. We also see that the posterior standard deviation of the spatial effect is lower in the vicinity of the 400 enumeration areas and higher elsewhere. The posterior median and 2.5th and 97.5th percentiles of the predicted average household wealth index is depicted in Figure \ref{Fig:kenyacomp_pred} of Appendix \ref{Ap:kenyasim2}.

\begin{table}[tbp]
\caption{Posterior median and 95\% credible intervals (CIs) for  parameters, mean squared error (MSE) and mean absolute error (MAE) of the surfaces in the simulation under five scenarios: 400 surveys with exact location (Surveys), census data at the county level (47 Areas), both survey and census data at the county level (Surveys + 47 Areas), census data at the provincial level (8 Areas), and both survey and census data at the provincial level (Surveys + 8 Areas).
\label{Table:sim}}
{\begin{center}
\small
\begin{tabular}{lcccccc}
\tblhead{Scenario & $\beta_0$: $0$ & $\sigma^2$: $0.25$ & $\phi$: $1.72$ & $\lambda^2$: $0.75$  & MSE & MAE}
\multirow{2}{*}{Surveys} & -0.0944  & 0.266 & 1.71 & 0.743  & 0.263 & 0.379 \\
& (-0.544, 0.371) & (0.213, 0.335) & (1.31, 2.32) & (0.474, 1.21) & & \\
\multirow{2}{*}{47 Areas} & -0.0307  & 0.210 & 1.80 & 0.712  & 0.307 & 0.431 \\
& (-0.494, 0.464) & (0.0620, 0.351) & (1.23, 2.64) & (0.429, 1.23) & & \\
\multirow{2}{*}{Surveys + 47 Areas} & -0.152  & 0.303  & 1.71 & 0.702 & 0.202 & 0.349 \\
& (-0.583, 0.264) & (0.245, 0.379) & (1.33, 2.27) & (0.464, 1.10) & & \\
\multirow{2}{*}{8 Areas} & -0.340  & 0.215  & 3.64 & 0.614 & 0.557 & 0.587 \\
& (-1.34, 0.863) & (0.0624, 0.362) & (1.40, 8.52) & (0.210, 2.16) & & \\
\multirow{2}{*}{Surveys + 8 Areas} & -0.135  & 0.318  & 1.66 & 0.688 & 0.254 & 0.386 \\
& (-0.558, 0.285) & (0.255, 0.399) & (1.27, 2.22) & (0.446, 1.10) & &
\lastline
\end{tabular}
\end{center}}
\end{table}

\subsection{Census Data (47 Counties)}
\label{sec:sim:area}

\begin{figure}[tbp]
\centering\includegraphics[width=.6\linewidth]{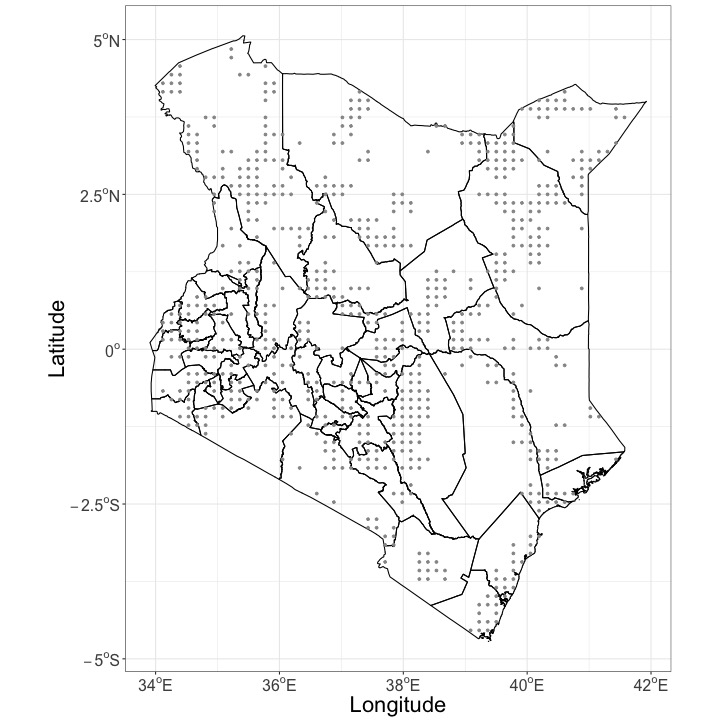}
\caption{Distribution of population in the 47 counties of Kenya. The gray circles represent mesh points where the population density is larger than average for that area.}
\label{Fig:kenyapop}
\end{figure}

We next consider a situation in which we have census data for each of the $n=47$ counties in Kenya. 
To implement (\ref{eq:censusmod}) we  approximate $\mu_i$ using (\ref{eq:normA}), which requires the population density at the mesh points. To determine the population estimate corresponding to the grid containing the mesh point, we used gridded population estimates from SEDAC.
Figure \ref{Fig:kenyapop} depicts the $n=47$ counties and mesh points with population density $d_{ik} > 1/m_i$ in gray.

The results are presented in Table \ref{Table:sim} and depicted in Figure \ref{Fig:kenyacomp} (second column). 
Again, we see that the posterior medians are relatively close to the true values. The predicted spatial surface is similar to the truth and is very similar to the predicted surface estimated for the point data. In general, the posterior standard deviation of the spatial effect is higher under this scenario than when we had information from 400 surveys. This is also evident when comparing the 2.5th and 97.5th percentile of the predicted average household wealth index, displayed in Figure \ref{Fig:kenyacomp_pred} of Appendix \ref{Ap:kenyasim2}.

\subsection{Survey and Census Data (47 Counties)}
\label{sec:sim:combo}

Another scenario that might arise is one in which we have both survey data at 400 EAs and census data available for 47 counties. Thus, we simply combine the methods from Sections \ref{sec:sim:points} and \ref{sec:sim:area}. The results are presented in Table \ref{Table:sim} and displayed in Figure \ref{Fig:kenyacomp} (third column) and Figure \ref{Fig:kenyacomp_pred} of Appendix \ref{Ap:kenyasim2}. Overall, there is a slight improvement over the survey information only case. We note that there are some identifiability problems when estimating the two variance parameters, which manifests itself here with $\sigma^2$ being overestimated and $\lambda^2$ underestimated.

\subsection{Census Data (8 Provinces)}

In order to evaluate the effect when area information is known at a more aggregate level than previously considered, we examined a situation where we only have census data available for each of the $n=8$ provinces in Kenya. Implementation-wise, this scenario is analogous to the one previously described in Section \ref{sec:sim:area}. Results are presented in Table \ref{Table:sim} and a depiction of the posterior mean and standard deviation along with a map displaying the 8 provinces is in Figure \ref{Fig:kenyacomp} (fourth column). In this scenario, inference for the parameters is severely deteriorated when compared to the previous cases. In particular, the credible intervals are much wider than in the previous scenarios and the MSE and MAE are substantially larger.

\subsection{Survey and Census Data (8 Provinces)}

The last scenario we consider is similar to that in Section \ref{sec:sim:combo}, where we have survey and census data available (at the provincial level) to use. Parameter estimates are presented in Table \ref{Table:sim} and the posterior mean and standard deviation of the random effect is depicted in Figure \ref{Fig:kenyacomp} (last column). Again, identifiability issues are evident in inference for the variances. The spatial effect surface is similar to the surveys-only scenario.

In terms of the mean squared errors, the values are 0.263, 0.307, and 0.202 when we have survey data with geographic coordinates, census data at the county-level ($n=47$), and a combination. In this simulation, there is a loss of accuracy when we only have census data, but it is not dramatic. However, when we aggregate at the provincial-level ($n=8$) the MSE is 0.557 and when we additionally incorporate survey data the MSE is 0.254. In general, we see a modest improvement when incorporating the census data over just using survey data. The improvement is significantly better when we used county-level census data rather than provincial-level census data. The same trends hold for the mean absolute errors.

\section{Application to Scottish Lip Cancer Data}
\label{sec:scotland}

We use the Scotland lip cancer data as an illustrative example of how the method can be applied to areal data.
The most common model for spatial smoothing for such data is that of \cite{besag:etal:91}. They propose a discrete spatial model by assigning the spatial random effects an intrinsic conditional autoregressive (ICAR) prior,
\[S_i~|~S_{i'},~ i' \in \text{ne}(i) \sim \text{N}\left(\overline{S}_i,\frac{1}{\tau_s m_i}\right)\]
where $S_{i'},~ i' \in \text{ne}(i)$ are the spatial random effects of the neighbors of $R_i$, $\overline{S}_i$ is the mean spatial random effect of the neighbors, $\tau_s$ determines the spatial variability, and $m_i$ is the number of neighbors. Unfortunately, this specification for the random effects depends on defining a somewhat arbitrary neighborhood structure.

Instead, it may be of interest to predict a continuously-varying rather than a discretely-varying latent surface. One may view the underlying continuous spatial field simply as a mechanism to induce spatial dependence between the areas (and then in some instances one may report the aggregate estimates only). One simplistic approach is to use a GRF model, with the data assumed to arise  (for example) at the centroids of the areas, but obviously this is arbitrary and does not reflect reality.

Let $R_i$ denote county $i$, $i=1,\dots,n=56$ and let $Y_{iaj}$ be the binary male lip cancer  indicator in stratum (age-band) $a$ of county $i$ at location $\bx_{ij}$, $j=1,\dots,N_{ia}$ where $N_{ia}$ is the male population in county $R_i$ age group $a$.
In the usual case, the available data correspond to summed disease counts $Y_{i++}=\sum_{a=1}^A\sum_{j=1}^{N_{ia}} Y_{iaj}$ and expected numbers $E_i$; these expected numbers are often pre-calculated as $E_i = \sum_{a=1}^A N_{ia} q_a$, where $q_a$ is a reference risk for stratum $a$. The $q_a$ may be taken from a previous time period or calculated (via internal standardization) in advance. The rarity of many diseases, and the lack of stratum-specific information, means that simplifying modeling assumptions are needed, as we now describe.

\begin{figure}[tbp]
\centering\includegraphics[width=0.85\linewidth]{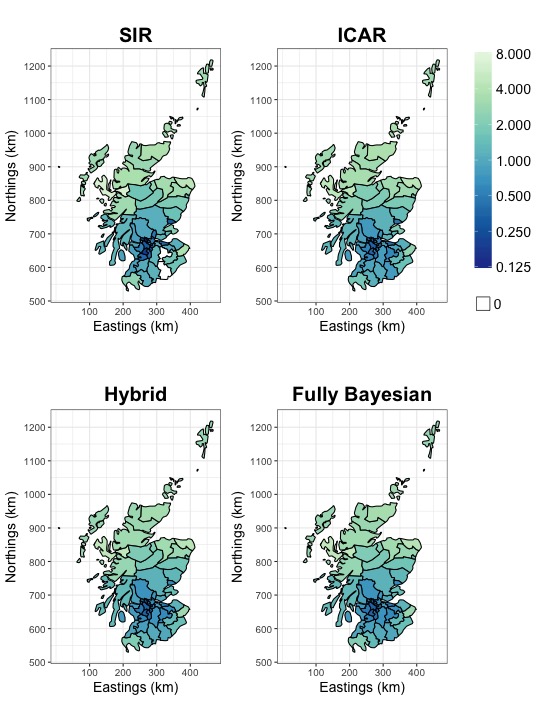}
\caption{Top left: the SIR estimates of the relative risks of lip cancer in 56 counties of Scotland. Top right: relative risk estimates (posterior medians) from the ICAR model. Bottom left: relative risk estimates (posterior medians) from aggregating results from the hybrid EB/MCMC approach. Bottom right: relative risk estimates (posterior medians) from aggregating results from the fully Bayesian approach.}
\label{Fig:scotsircomp}
\end{figure}

We proceed as in the no strata case and assume for a rare disease $Y_{iaj} ~|~ p_{iaj} \sim_{ind} \mbox{Poisson}(p_{iaj})$, for $j=1,\dots,N_{ia}$ individuals in strata $a$, county $R_i$, where $p_{iaj}=\exp \{ \beta_0+\beta_a + S_a(\bx_{ij})\}= q_a \exp \{ \beta_0 + S_a(\bx_{ij})\},$
with $S_a(\bx_{ij})$ representing the spatial random effect for strata $a$ at location $\bx_{ij}$.
This leads to
$Y_{i++} | \mu_i \sim \mbox{Poisson}(\mu_i) $,
and, proceeding as before,
\begin{eqnarray*}
\mu_i &=& \sum_{a=1}^A\sum_{j=1}^{N_{ia}} \E_a
\left[ Y_{iaj} | \bx_{ij}\right] = \sum_{a=1}^A q_{a} \sum_{j=1}^{N_{ia}} \exp\{\beta_0 + S_a(\bx_{ij})\}\\
&\approx& \sum_{a=1}^A N_{ia} q_a  \exp(\beta_0) \sum_{k=1}^{m_{i}} d_{a}(\bx_{ik})\exp \{ S_a(\bx_{ik}) \} \\&=& E_i \exp(\beta_0) \sum_{k=1}^{m_{i}} d(\bx_{ik})\exp \{ S(\bx_{ik}) = E_i \theta_i
\end{eqnarray*}
where the first equality on the last line follows from assuming a common residual spatial risk surface across stratum ($S_a(\bx) = S(\bx)$) and common population density across stratum ($d_{a}(\bx) = d(\bx)$). This allows us to separate the age-standardization from the risk surface estimation to give the data model.
Standardization in this fashion leads to the spatial modeling of the {\it relative risk}, $\theta_i$, an aggregate summary.
The standardized incidence ratio (SIR) is  $\mbox{SIR}_i=Y_i/E_i$ and is the MLE of $\theta_i$ from the Poisson model with mean $E_i\theta_i$. The SIRs are depicted in the top left hand panel of Figure \ref{Fig:scotsircomp}. It is evident from the map that there is large variability in the area relative risks, with apparent strong spatial dependence.

\begin{figure}[tbp]
\centering\includegraphics[width=1\linewidth]{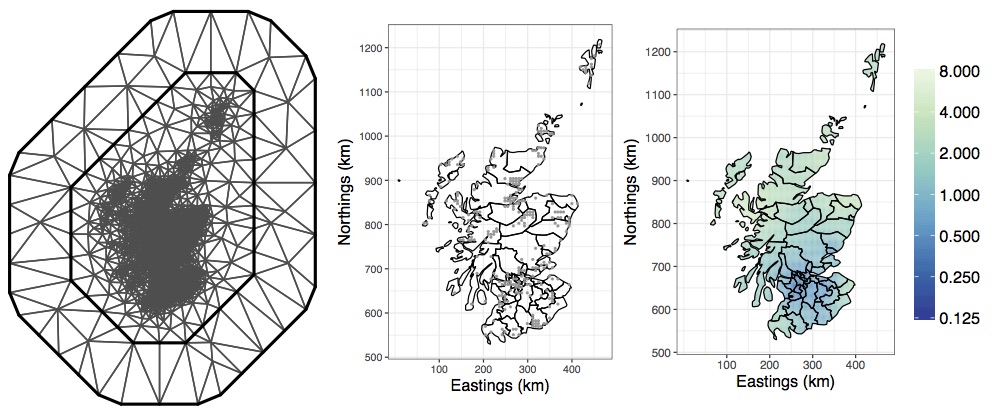}
\caption{Left: mesh used for Scotland analysis, consisting of $m=2,417$ mesh points. Middle: distribution of population in Scotland. The gray circles represent mesh points where the population density is larger than average for that area. Right: the predicted continuous relative risk surface from using the fully Bayesian approach.}
\label{Fig:scotsummary}
\end{figure}

Inference for this model proceeds as discussed in Section \ref{sec:poissoncomp}.
The mesh used is shown in Figure \ref{Fig:scotsummary} with $m=2,417$ mesh points, which results in mesh points that are $\approx$ 6.3\ km apart. We determined the relative population density for each $R_i$ in the same manner as we did for the Kenya simulation and mesh points associated with higher relative population densities are shown in Figure \ref{Fig:scotsummary}.

We considered several different computational strategies, which are described more fully in Appendix \ref{Ap:scotcomp}. Briefly, we implemented an EB approach in which the spatial random effects were integrated out using Laplace approximations, a fully Bayesian approach (using HMC), and a hybrid of these two in which HMC was used, with $\btheta$ fixed at the EB estimates. For the fully Bayesian approach, we initialized 4 chains, used a burn in of 10,000 iterations, ran up to an additional 1,000,000 iterations and thinned them to ultimately save 1,000 iterations from each chain. For the hybrid approach, we also initialized 4 chains, used a burn in of 500 iterations, ran up to an additional 1,000 iterations for each chain.
Convergence summaries for both the fully Bayesian and hybrid approach are given in Appendix \ref{Ap:scotconv}. 

Estimates and 95\% CIs for the parameters $\exp(\beta_0)$, $\phi$, and $\lambda^2$ are presented in Table \ref{Table:scot} for the three different computational strategies. There is good agreement, though we notice that the posterior credible intervals tend to be wider when using the fully Bayesian computational strategy, which is not surprising given the use of the delta-method to calculate the CIs in the EB approach.

\begin{table}[tbp]
\caption{Estimates and 95\% credible intervals (CIs) for parameters in the Scotland example under the three different computational approaches. For the empirical Bayes approach, estimates are based on transformed MLEs and CIs are based on the delta-method. For the hybrid approach, estimates are based on transformed MLEs for the spatial parameters and the posterior median and 2.5th and 97.5th percentile are presented for the intercept. For the fully Bayesian approach, estimates reported are posterior medians and CIs are based on the 2.5th and 97.5th percentiles.
\label{Table:scot}}
{
\begin{center}
\begin{tabular}{cccc}
\tblhead{
Parameter & Empirical Bayes & Hybrid & Fully Bayesian }
$\exp(\beta_0)$ & 1.99 (1.35, 2.94) & 2.03 (1.36, 3.01) & 1.91 (1.26, 3.13)  \\
$\phi$ & 71.8 (35.1, 147) & 80.3 (38.9, 166) &  85.3 (42.0, 203)  \\
$\lambda^2$ & 0.516 (0.279, 0.952) & 0.534 (0.283, 1.01)  & 0.581 (0.297, 1.28) 
\lastline
\end{tabular}
\end{center}}
\end{table}

We also obtain predictions and posterior standard deviations of $\tilde{S}(\bx)$, displayed in Figure \ref{Fig:scotfieldpred} of Appendix \ref{Ap:scotconv}. We note that the posterior standard deviation of the surface is smallest in regions of Scotland where the population is greatest, and larger elsewhere. Furthermore, the posterior standard deviation tends to be a little lower for the hybrid approach than for the fully Bayesian approach, which is not surprising given that the spatial parameters were fixed in the hybrid approach. The predicted continuous relative risk surface using the fully Bayesian approach (posterior median) is presented in Figure \ref{Fig:scotsummary}. We see that the continuous relative risk surface and is largest in the counties with higher SIRs, and lowest in the counties with the smallest SIRs.

We obtain relative risk estimates (posterior medians), as well as 95\% CIs for each of the 56 counties from this model by aggregating the continuous relative risk surface within each county; posterior medians are presented in Figure \ref{Fig:scotsircomp} and the 95\% CIs are displayed in \ref{Fig:scotprevcis} of Appendix \ref{Ap:scotconv}. To obtain estimates of the desired quantiles in both the fully Bayesian and hybrid approach, for each county $R_i$ we obtain $b=1,\dots,B=4,000$ draws from the ``aggregated'' relative risk surface, $\theta_i^{(b)} = \exp\left(\beta_0^{(b)}\right) \ \bD_i^\text{T} \bT^{(b)}$, where $\bT^{(b)} = \left[~\exp\left(w_1^{(b)}\right),\dots,\exp\left(w_m^{(b)}\right)~ \right]^\text{T}$ (see (\ref{eq:poisA})). As before, $\bD_i$ has at most $m_i$ nonzero entries that correspond to the population density estimates, $d(\bx_{ik})$. From here, we can obtain the desired summary measures. We see that the relative risk estimates are nearly identical for both computational strategies and are similar to the SIRs, but that the estimates are shrunk towards the overall mean, which is as expected.

We also compare our results to those obtained using an ICAR prior on the spatial random effects. Parameter estimates are in Appendix \ref{Ap:scotconv}.  
The predicted relative risks (posterior medians) for each county are presented in Figure \ref{Fig:scotsircomp}, and the 95\% CIs are presented in Figure \ref{Fig:scotprevcis} of Appendix \ref{Ap:scotconv}.
The results are very similar to the continuous model. 

For the fully Bayesian approach using HMC (using our own code), it took approximately a week to fit the model using a computing cluster. This can be improved tremendously by using the hybrid approach. It takes on the order of minutes to obtain the empirical Bayes estimates and about ten minutes to run the HMC.

\section{Discussion}
\label{sec:discussion}

In this article, we propose a Bayesian hierarchical model that can accommodate observations taken at different spatial resolutions. To this end, we assume a continuous spatial surface, which we model using the SPDE approach.

In the simulation example, we considered surveys taken at point locations and census data associated with areas. 
When the only data available was census data at the county-level, there was not a substantial loss in accuracy when comparing it to a situation in which we had survey (point) data.
In general, we would not expect this to hold when comparing point data to areal data as the loss of information depends on both the strength of the spatial dependence, the number and geographical configuration of the areas, and on the amount and quality of the survey data. When the size of the areas increased (comparing 47 counties to 8 provinces), estimates for the spatial parameters and the overall household wealth index were highly variable and the predicted spatial surface was much less nuanced.

We also applied our method to the Scotland lip cancer dataset. Overall, there were very minor differences in the relative risk estimates for each county when comparing our continuous spatial model to a discrete spatial model (i.e.,~the ICAR model) but again there is strong spatial dependence in these data (which explains in part the popularity of these data). However, we note that modeling a continuous surface is particularly attractive in that it is not subject to definitions of administrative boundaries, which can often be arbitrary, and a continuous risk surface, in general, more accurately reflects disease etiology. Furthermore, it can easily be adapted to situations where we might also have point level covariate data. In the latter case, the use of the models we have described avoids the ecological fallacy which occurs when area-level associations differ from the individual-level counterparts. To avoid ecological bias, one requires point level covariates but the availability of such data is increasing \citep{gething:etal:15}.

For normal outcomes, all computation can be performed quickly using \texttt{R-INLA}. In the simulation example, it took about 2 minutes to fit each of the models on a standard laptop. For Poisson outcomes, computation is much more difficult.  There has been an increased interest in implementing sparse matrix operations in Stan, which would improve usability of this method. In general, there is still a need for computationally efficient MCMC schemes for Gaussian process data.

With point data we can't learn anything about the surface at a spatial resolution which is less than the distance between the two closest points. If we only have areal data, the situation is far worse. Hence, one should not  over-interpret fine spatial scale effects. In both the point and area data cases, model checking is difficult.

As a minimum however, for areal data, one may simply view the model as a method to induce an area-level spatial prior. If there are doubts on the fine-scale surface, the results can be presented at the area-level, as we did with the Scottish lip cancer example.

In the simulation study we considered, we looked at combining survey (point) data and census (areal) data. However, with older DHS surveys exact geographic coordinates are not available. Instead, it is only known in which area of the country the survey was taken. This is different from the problem we considered in that, instead of observing outcomes that are associated with an entire area (census data), outcomes are for a specific point in the area, but that exact location is unknown. Therefore, the methods proposed here would need to be altered to accommodate this type of situation.

\section{Software}

All code and input data used in the simulation and application is available on github (\url{https://github.com/wilsonka/pointless-spatial-modeling}).

\section*{Acknowledgments}

The authors would like to thank Dan Simpson for numerous helpful conversations, and Jim Thorson for advice on TMB. Both authors were supported by R01CA095994 from the National Institutes of Health.

\bibliographystyle{apalike}
\bibliography{main}

\clearpage

\section*{Appendix}
\appendix
\label{appendix}

\counterwithin{figure}{section}

\section{Computational Details for the Kenya Simulation}
\label{Ap:kenyasim1}

As described in Section 4.1 of the paper, the household-level model is
$Y_{ih} ~|~ \mu_{ih} \sim \N(\mu_{ih},\sigma^2)$, with $$\mu_{ih} = \beta_0 + S(\bx_{ih}),$$
with unknown variance $\sigma^2$. 
Across all simulation scenarios, we use the priors, $\beta_0  \sim \N(\mu_{\beta_0},\sigma_{\beta_0}^2)$, $\sigma^2 \sim \text{Beta}(2, 5)$, 
$\btheta  \sim \N\left(\bmu_{\theta}, \bSigma_{\theta}\right)$,
with $\mu_{\beta_0}=0$, $\sigma_{\beta_0}^2=100$, 
$$\bmu_{\theta}=\left[\begin{array}{c}
-1.17 \\
-0.0933
\end{array} \right],\qquad \bSigma_{\theta}=\left[\begin{array}{cc}
10 & 0 \\
0 & 10
\end{array} \right].$$
The hyperprior for $\btheta$ is chosen to be fairly vague. Here, the prior mean for $\theta_1$ corresponds to a marginal variance $\lambda^2$ of 1. 
The prior mean for $\theta_2$ corresponds to a practical range $\phi$ of roughly 20\% of the domain size.

We use the \texttt{R} package \texttt{R-INLA} for computation. Fitting models involving observations with exact locations is straightforward as there exist functions to define the matrix $\bA$ used to project the spatial random effect from the mesh vertices to point locations (see Section 3.1). 
Details of how to specify these models in \texttt{R-INLA} using the SPDE approach can be found in \cite{lindgren:rue:15}.
In order to fit the models that involve census data, we adapt $\bA$ since this matrix can be viewed as a way to average the random effect at mesh points.
In these scenarios, we define $\bD$ to be a matrix with $n$ rows and $m$ columns, made up of row-vectors $\bD_{i}^\text{T}$ of length $m$, where $m$ is the number of mesh points. These row vectors $\bD_{i}^\text{T}$ contain up to $m_i$ non-zero entries $d(\bx_{ik})$. In the case of area-level observations, we use $\bD$ in place of $\bA$ when fitting the model using \texttt{R-INLA}. In scenarios involving a combination of point and areal data, the resulting projection matrix contains rows from both $\bD$ and $\bA$.

\section{Further Material on the Kenya Simulation}
\label{Ap:kenyasim2}

\begin{figure}[tbp]
\centering\includegraphics[width=0.98\linewidth]{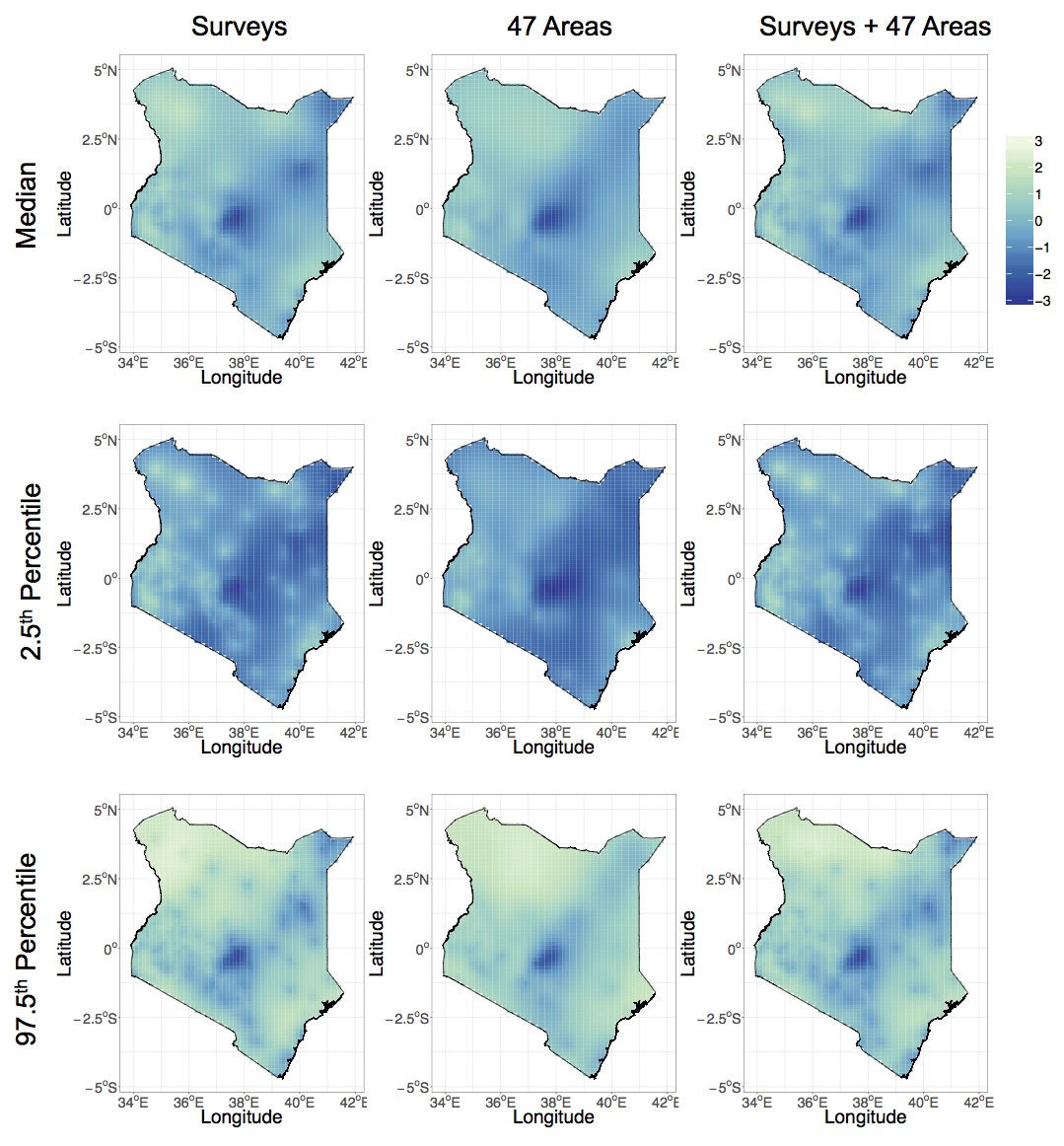}
\caption{Comparison of results when the 400 surveys are used (Left), census data from the 47 counties are used (Middle), and both are used (Right). Top row is the predicted (posterior median) household wealth index surface, middle row is the 2.5 percentile, and bottom row is the 97.5 percentile.}
\label{Fig:kenyacomp_pred}
\end{figure}

We compare the predicted average household wealth index (posterior median) and 2.5th and 97.5th percentiles across three scenarios: 400 surveys with exact location, census data at the county level, and both survey and census data at the county level. Results are presented in Figure \ref{Fig:kenyacomp_pred}. When comparing to the true household wealth index surface, which tends to have lower values in the middle and northeastern sections of the country and higher values elsewhere, we see that the predicted surface when using the survey coordinates (``Surveys'' and ``Surveys + 47 Areas'') tends to be most similar.

\section{Computational Details for the Scotland Example}
\label{Ap:scotcomp}

In this section, we describe computational strategies for analyzing the Scotland dataset. We first describe two empirical Bayes based approaches followed by a description of the fully Bayesian approach.

\subsection{Empirical Bayes}

As fast alternatives to a completely Bayesian approach, we consider two strategies, both based on empirical Bayes (EB) estimation. In the first, we use EB estimation to obtain estimates for the spatial hyperparameters $\btheta$ and the fixed effect $\beta_0$. In the second, we use a hybrid approach where we first use EB estimation to estimate $\btheta$, and then proceed, conditional on these values.

For the first, strictly EB, approach, the EB estimates are defined as 
\begin{align*}
\left(\hat{\btheta}^\text{EB},~ \hat{\beta}_0^\text{EB} \right) & = \mbox{argmax}_{\theta, \beta_0}\ p(\btheta, \beta_0|\by)\\
& = \mbox{argmax}_{\theta, \beta_0} \int_{w} p(\btheta,\bw,\beta_0|\by)~ d\bw,\\
& = \mbox{argmax}_{\theta, \beta_0} \int_{w} f(\by|\beta_0,\bw)p(\bw|\btheta)\tilde{p}(\btheta)\tilde{p}(\beta_0)~d\bw,
\end{align*}
where we use $\tilde{p}(\cdot)$ to denote a flat prior, that is $\tilde{p}(\cdot) \propto 1$. For the EB approach we use uninformative, flat priors for both the fixed  effect $\beta_0$ and spatial parameters $\btheta$. Therefore, the EB estimates, the posterior modes, are also maximum likelihood estimates (MLEs). We then use the invariance of MLEs and the delta-method to obtain estimates for $\exp(\beta_0)$ and functions of the spatial hyperparameters. 

For the second, ``hybrid'', approach, the EB estimates are defined as
\begin{align*}
\hat{\btheta}^\text{Hybrid} & = \mbox{argmax}_{\theta}\ p(\btheta|\by)\\
& = \mbox{argmax}_{\theta} \int_{w}\int_{\beta_0} p(\btheta,\bw,\beta_0|\by)~d\beta_0~ d\bw,\\
& = \mbox{argmax}_{\theta} \int_{w}\int_{\beta_0} f(\by|\beta_0,\bw)p(\bw|\btheta)\tilde{p}(\btheta)p(\beta_0)~d\beta_0~ d\bw.
\end{align*} 
We then use these estimates in the second, MCMC-based, step, which is described in the following section.
In the hybrid approach, we use a normal prior for the intercept, $\beta_0 \sim \N(\mu_{\beta_0}, \sigma_{\beta_0}^2)$  with $\mu_{\beta_0}=0,\sigma_{\beta_0}^2=100$, and place uninformative, flat priors on the spatial parameters, $\tilde{p}(\btheta) \propto 1$.

In both cases, to implement EB estimation we use the \texttt{R} package \texttt{TMB} \citep{kristensen:14,kristensen:15}, which is computationally efficient since it uses sparse matrix operators. See \cite{thorson:15} for a discussion on using \texttt{TMB} with the SPDE approach.

We briefly summarize how to implement this approach. We first construct a so-called template file that contains the joint distribution, which is the product of the likelihood $f(\by|\beta_0, \bw)$ and priors $p(\bw|\btheta)$, $\tilde{p}(\beta_0)$ or $p(\beta_0)$, and $\tilde{p}(\btheta)$. To obtain the densities that we would like to optimize, $p(\btheta, \beta_0 | \by)$ and $p(\btheta|\by)$, we use {\tt TMB} to integrate out $\bw$ and, optionally, $\beta_0$. This integration is carried out using Laplace approximations. We then numerically optimize the density using gradients to obtain the EB estimates denoted $\hat{\btheta}^{\text{EB}}$ and $\hat{\beta}_0^{\text{EB}}$ (or $\hat{\btheta}^\text{Hybrid}$ for the hybrid approach) and the associated variance-covariance matrix (based on the Hessian), denoted $\hat{\bSigma}^\text{EB}$ (for the strictly EB approach).

\subsection{MCMC}
For the fully Bayesian approach we use as priors,
$\beta_0  \sim \N(\mu_{\beta_0},\sigma_{\beta_0}^2)$ and
$\btheta  \sim \N\left(\bmu_{\theta}, \bSigma_{\theta}\right)$,
with $\mu_{\beta_0}=0,\sigma_{\beta_0}^2=100$,
$$\bmu_{\theta}=\left[\begin{array}{c}
3.24 \\
-4.51
\end{array} \right],\qquad \bSigma_{\theta}=\left[\begin{array}{cc}
10 & 0 \\
0 & 10
\end{array} \right].$$
As in the Kenya simulation example, the priors for the hyperparameters $\btheta$ are vague and $\bmu_{\theta}$ corresponds to a marginal variance $\lambda^2$ of 1 and practical range $\phi$ of roughly 20\% of the domain size.

In the fully Bayesian approach, we first begin by updating $\btheta$ conditional on $\bw$, $\beta_0$, and $\by$ using a random-walk proposal. The proposal distribution we use is,
\begin{align*}
\btheta^{(t+1)} \sim \N(\btheta^{(t)},c\times \hat{\bSigma}^\text{EB}_{\theta}),
\end{align*}
where $\hat{\bSigma}^\text{EB}_{\theta}$ is the inverse Hessian corresponding to the estimates for $\btheta$ obtained from EB estimation described in the preceding section.

The second step is then similar for both the hybrid and fully Bayesian approach. We update $\bw$ and $\beta_0$ conditional on $\btheta$ and $\by$, by using Hamiltonian Monte Carlo (HMC; Neal, 2011\nocite{neal:2011}). In the hybrid approach, $\btheta$ is taken to be $\hat{\btheta}^{\text{Hybrid}}$. The negative log posterior $U$ (modulo a constant term), is found to be
\begin{align*}
U =  -\beta_0 \by^\text{T}\mathbf{1}_n - \by^\text{T} \log(\bD \bT) + \exp(\beta_0)\bE^\text{T}\bD \bT + \frac{1}{2}\bw^\text{T}\bQ\bw + \frac{1}{2\sigma_{\beta_0}^2}\beta_0^2,
\end{align*}
where $\mathbf{1}_n$ is an $n \times 1$ vector of all ones, $\bD$ is an $n \times m$ matrix where each row $\bD_i^\text{T}$ contains up to $m_i$ nonzero weights $d(\bx_{ik})$,  $\bT=[~\exp(w_1),\dots,\exp(w_m)~]^\text{T}$, and $\bE = [~E_1,\dots,E_n~]^\text{T}$. We also compute the derivatives to be,
\begin{align*}
\frac{\partial U}{\partial \beta_0} & = -\by^\text{T}\mathbf{1}_n + \exp(\beta_0) \bE^\text{T} \bD \bT + \frac{\beta_0}{\sigma_{\beta_0}^2},\\
\frac{\partial U}{\partial \bw} & = \left(\bD \ \text{diag}(\bT)\right)^\text{T} \left[ \exp(\beta_0)\bE - \left(\text{diag}(\bD \bT) \right)^{-1} \by\right] + \bw^\text{T}\bQ,
\end{align*}
where $\text{diag}(\bT)$ is a diagonal matrix with entries $T_1,\dots,T_m$ along the diagonal.
Parameters that are tuned for desired acceptance are $c$, the step size, and number of leapfrog steps for each HMC iteration.

In exploratory runs, it appeared that computation could be improved by defining $\bw^* = \bw + \beta_0$, in which case, $\bw^*~|~\beta_0,\btheta \sim N(\beta_0, \mathbf{Q}^{-1})$. Under this parameterization,
\begin{align*}
U & =  - \by^\text{T} \log(\bD \bT^*) + \bE^\text{T}\bD \bT^* + \frac{1}{2}(\bw^* - \beta_0 \mathbf{1}_m)^\text{T}\bQ(\bw^* - \beta_0 \mathbf{1}_m) + \frac{1}{2\sigma_{\beta_0}^2}\beta_0^2,\\
\frac{\partial U}{\partial \beta_0} & = \beta_0 \mathbf{1}_m^\text{T}\bQ\mathbf{1}_m - \mathbf{1}_m^\text{T}\bQ\bw^* + \frac{\beta_0}{\sigma_{\beta_0}^2},\\
\frac{\partial U}{\partial \bw^*} & = \left(\bD \ \text{diag}(\bT^*)\right)^\text{T} \left[ \bE - \left(\text{diag}(\bD \bT^*) \right)^{-1} \by\right] + \bw^{*\text{T}}\bQ - \beta_0 \bQ \mathbf{1}_m,
\end{align*}
where $\bT^*=[~\exp(w^*_1),\dots,\exp(w^*_m)~]^\text{T}$. This alternative parameterization is implemented for the hybrid approach. 

Further gains in speed can be found by specifying a better scaling of the ``mass matrix'' (the covariance of the momentum variables used in the HMC algorithm). In the most simple case, the mass matrix  is chosen to be the identity matrix, which corresponds to i.i.d. momentum variables. For the hybrid approach, we adapt this matrix to again be diagonal, but with entries along the diagonal corresponding to the inverse posterior variance. We empirically estimate this after running several hundred iterations of the algorithm using an identity mass matrix.

\section{Further Material on the Scottish Lip Cancer Example}
\label{Ap:scotconv}

\begin{figure}[tbp]
\centering\includegraphics[width=0.85\linewidth]{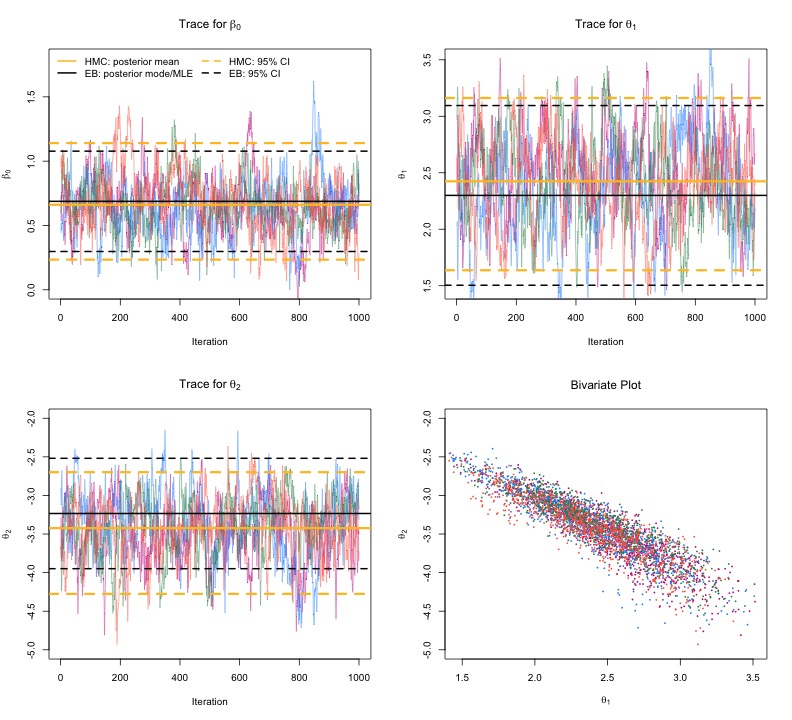}
\caption{Trace plots for $\beta_0$, $\theta_1$, and $\theta_2$ in the Scotland example using the fully Bayesian approach. Solid gold lines and dashed gold lines are the posterior means and 95\% CI, respectively using HMC. Solid black lines are the EB estimates and the dashed black lines are the corresponding 95\% CI using the strictly EB approach.}
\label{Fig:scottrace}
\end{figure}

\begin{figure}[tbp]
\centering\includegraphics[width=0.85\linewidth]{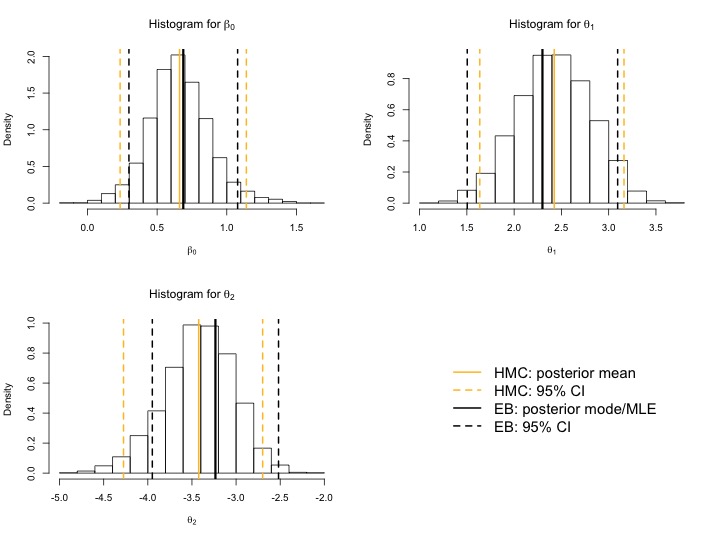}
\caption{Univariate posterior distributions for $\beta_0$, $\theta_1$, and $\theta_2$ using the fully Bayesian approach. Solid gold lines and dashed gold lines are the posterior means and 95\% CI, respectively using HMC. Solid black lines are the EB estimates and the dashed black lines are the corresponding 95\% CI using the strictly EB approach.}
\label{Fig:scothist}
\end{figure}

\begin{figure}[tbp]
\centering\includegraphics[width=0.85\linewidth]{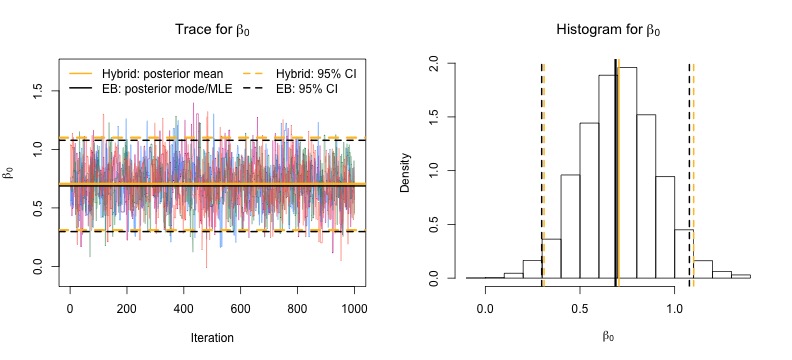}
\caption{Trace plot and univariate posterior distribution for $\beta_0$ in the Scotland example using the hybrid approach. Solid gold lines and dashed gold lines are the posterior means and 95\% CI, respectively. Solid black lines and dashed black lines are the EB estimates and 95\% CI, respectively using the strictly EB approach.}
\label{Fig:scotconv}
\end{figure}

Trace plots and histograms for the fully Bayesian approach are shown in Figures \ref{Fig:scottrace} and \ref{Fig:scothist}, respectively. The trace plot and histogram for the hybrid approach is displayed in Figure \ref{Fig:scotconv}. Trace plots and calculated $\hat{R}$ (which were all less than 1.05) suggested convergence \citep{gelman:2014}.

\begin{figure}[tbp]
\centering\includegraphics[width=0.85\linewidth]{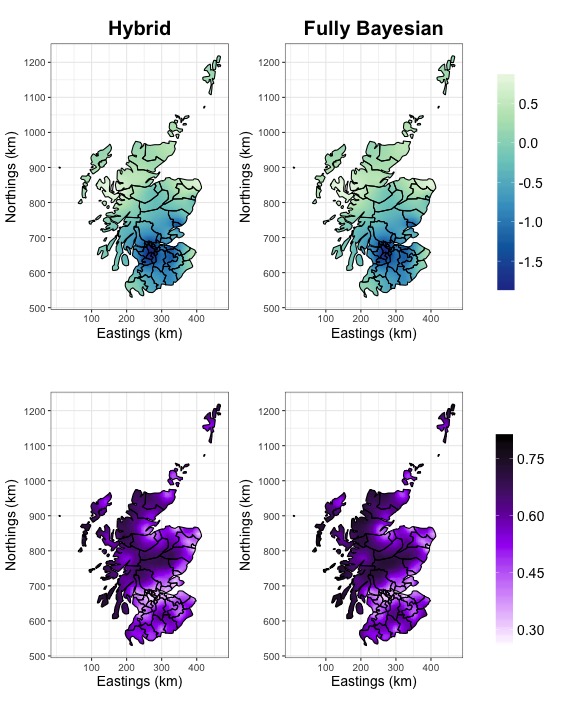}
\caption{Top: predicted (posterior mean) spatial surface. Bottom: posterior standard deviation of the spatial surface. Left column are results from the hybrid approach and right column are results from the fully Bayesian approach.}
\label{Fig:scotfieldpred}
\end{figure}

The predicted spatial surface (posterior mean) and corresponding posterior standard deviation are depicted in Figure \ref{Fig:scotfieldpred} for both the hybrid (left column) and fully Bayesian approach (right column). In both cases the predicted continuous surface is similar and the posterior standard deviation follow similar trends. However, the posterior standard deviation tends to be lower in the hybrid approach when compared to the fully Bayesian approach, which is as expected since variability in $\btheta$ is not taken into consideration in the hybrid approach.

Using the ICAR model, we obtained a posterior median for $\exp(\beta_0)$ of 1.03 (95\% CI: 0.926, 1.14) and for $\tau_s$ of 1.33 (95\% CI: 0.770, 2.28).

\begin{figure}[tbp]
\centering\includegraphics[width=1\linewidth]{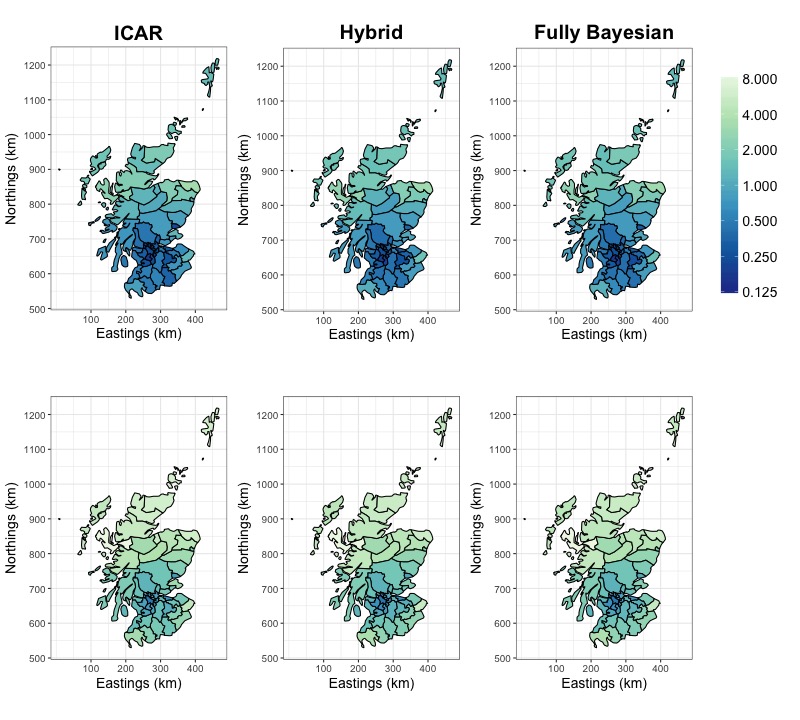}
\caption{Comparison of the 95\% CI of relative risk under different models for the spatial surface and computational strategies. Left column are estimates from the ICAR model, middle column are estimates from using the hybrid approach, and right column are estimates from the fully Bayesian approach. Top row is 2.5th percentile, and bottom row is 97.5th percentile.}
\label{Fig:scotprevcis}
\end{figure}

A comparison of the 95\% CIs for the relative risk using the ICAR model and continuous surface model are presented in Figure \ref{Fig:scotprevcis}. Plotted are the corresponding 2.5th percentiles and 97.5th percentiles. Results were nearly indistinguishable across the models and computational strategies.

\end{document}